\documentstyle[preprint,prb,aps]{revtex}

\begin{document}
\title{$s$- and $d_{xy}$-wave components
induced around a vortex \\
in $d_{x^2-y^2}$-wave superconductors}
\author{M. Ichioka}
\address{Department of Physics, Kyoto University,
         Kyoto 606-01, Japan}
\author{N. Enomoto, N. Hayashi and K. Machida}
\address{Department of Physics, Okayama University,
         Okayama 700, Japan}
\date{\today}
\maketitle
\begin{abstract}
 Vortex structure of $d_{x^2-y^2}$-wave superconductors
is microscopically analyzed in the framework of
the quasi-classical Eilenberger equations.
If the pairing interaction contains an $s$-wave ($d_{xy}$-wave)
component in addition to a $d_{x^2-y^2}$-wave component,
the $s$-wave ($d_{xy}$-wave) component of the order parameter
is necessarily induced around a vortex in $d_{x^2-y^2}$-wave
superconductors.
 The spatial distribution of the induced $s$-wave and $d_{xy}$-
wave
components is calculated.
 The $s$-wave component has opposite winding number around
vortex near the $d_{x^2-y^2}$-vortex core and its amplitude has
the shape of a four-lobe clover.
The amplitude of $d_{xy}$-component has the shape of an
octofoil.
    These are consistent with results based on the GL theory.
\end{abstract}
\narrowtext

In the last years, a number of investigations were
carried out theoretically and experimentally to identify
the symmetry
of pairing state in high-$T_c$ superconductors.
Although precise pairing symmetry has not been determined yet,
it is recognized that $d_{x^2-y^2}$-wave symmetry is most
probable.
\cite{m2s}
    Recently, the vortex structure of the $d$-wave
superconductors attracts much attention
because it may have different structure from that of
conventional $s$-wave superconductors.
    One of the related topics is a possibility that
other components of the order parameter may be induced
around a vortex in $d$-wave superconductors.
    Based on symmetry considerations,
Volovik suggested this possibility. \cite{vol}
    In $d_{x^2-y^2}$-wave superconductors,
the amplitude of $s$-wave pairing
should be contained at the core region of the vortex,
since the $d_{x^2-y^2}$-wave vortex has the same symmetry
as that of the opposite winding $s$-wave component.
    Solving the tight-binding Bogoliubov-de Gennes equation
self-consistently,  Soininen, Kallin and Berlinsky calculated the
vortex structure
on a 16$\times$16 lattice
and showed that $s$-wave component with an opposite winding
number
is contained around the vortex core in $d_{x^2-y^2}$-wave
superconductors. \cite{soi}
    Ren, Xu and Ting derived the two components
Ginzburg Landau (GL) equations for $s$- and $d$-wave
superconductivity from the Gor'kov equations,
and studied the mixing of the $s$-wave component
near a vortex in $d$-wave superconductors. \cite{ren}
    They suggested that the winding number of the induced
$s$-wave component around the vortex is 3 far from
the vortex core and $-1$ near the center of vortex,
and its profile has the shape of a four-lobe clover.

According to the general consideration based on the GL theory,
it is possible that the $s$-wave component is coupled with
the $d$-wave component through the gradient terms.
Therefore, $s$-wave component is induced wherever
the $d$-wave order parameter spatially varies,
such as near the vortex.
    The mixing due to the same mechanism also occurs
near a surface of $d$-wave superconductors,
where $d$-wave component decreases on approaching the surface
and $s$-wave component is induced in the surface region.
    This surface effect was considered
by Matsumoto and Shiba using the self-consistent
quasi-classical Green function formalism.\cite{mat}
    They also considered the mixing of $d_{xy}$-wave
component in addition to the mixing of $s$-wave component.

    The purpose of this paper is to analyze $s$- and
$d_{xy}$-wave components induced around a vortex in
$d_{x^2-y^2}$-wave superconductors using the quasi-classical
Eilenberger equations, \cite{eil}
which can be applied at arbitrary temperatures.
    The quasi-classical calculations on the vortex structure
were carried out for conventional $s$-wave superconductors
by Pesch and Kramer,\cite{pes} and Klein,\cite{kle}
and for pure $d_{x^2-y^2}$-wave superconductors
by Schopohl and Maki,\cite{sch,mak}
and the current authors.\cite{ich}
 Here we consider the case of an isolated vortex
 under a magnetic field applied parallel
to the $c$-axis (or $z$-axis) in the clean limit.
    The Fermi surface is assumed to be two-dimensional,
which is appropriate to high-$T_c$ superconductors,
and isotropic for simplicity.
    Throughout the paper, lengths and energies are measured
in units of the coherence length $\xi$ and
the uniform gap $\Delta_0$ at $T=0$, respectively.

First, we solve the quasi-classical Eilenberger equations to obtain
the Green functions for a given pair potential.\cite{self}
For the $d_{x^2-y^2}$ symmetry and not too low temperatures,
the pair potential may be assumed by
\begin{eqnarray}
&&
\Delta(\theta,{\bf r})
= \bar\Delta(\theta,{\bf r}) e^{i\phi} ,\nonumber \\
&&
\bar\Delta(\theta,{\bf r})
= \Delta(T)\tanh(r)\cos(2\theta) ,
\label{eq:1}
\end{eqnarray}
where $r=\sqrt{x^2+y^2}$ is the distance
from the center of vortex line,
$\theta$ is the angle of ${\bf k}$-vector
with the $a$-axis (or $x$-axis),
and the phase $\phi$ of the pair potential around vortex center
is given by $e^{i\phi} =(x+iy)/r$.
The quasi-classical Green functions with the
Matsubara frequency $\omega_n=(2n+1)\pi T$
are obtained
by solving the Eilenberger equations,
which are given as follows in the gauge
where pair potential is real,
\begin{equation}
\Bigl\{ \omega_n +{1 \over 2}
\Bigl(\partial_\parallel
+ i\partial_\parallel \phi \Bigr) \Bigr\}{\bar f}
(\omega_n,\theta,{\bf r})
= \bar\Delta(\theta,{\bf r}) g(\omega_n,\theta,{\bf r}) ,
\label{eq:2}
\end{equation}
\begin{equation}
\Bigl\{ \omega_n -{1 \over 2}
\Bigl(\partial_\parallel
-i\partial_\parallel \phi \Bigr) \Bigr\}
{\bar f}^\dagger(\omega_n,\theta,{\bf r})
= \bar\Delta(\theta,{\bf r}) g(\omega_n,\theta,{\bf r}),
\label{eq:3}
\end{equation}
\begin{equation}
g(\omega_n,\theta,{\bf r})
=\Bigl(1-{\bar f}(\omega_n,\theta,{\bf r})
    {\bar f}^\dagger(\omega_n,\theta,{\bf r}) \Bigr)^{1/2},
\label{eq:4}
\end{equation}
where $ {\rm Re} g(\omega_n,\theta,{\bf r}) > 0 $ and
$\partial_\parallel \phi= -r_\perp /r^2$.
    Here, we have taken the coordinate system:
$\hat{\bf u}=\cos\theta \hat{\bf x}+\sin\theta \hat{\bf y}$,
$\hat{\bf v}=-\sin\theta \hat{\bf x}+\cos\theta \hat{\bf y}$,
thus a point ${\bf r}=x \hat{\bf x}+y \hat{\bf y}$
is denoted as
${\bf r}=r_\parallel \hat{\bf u}+r_\perp \hat{\bf v}$.
    The anomalous Green functions
${\bar f}$ and ${\bar f}^\dagger$
in Eqs.(\ref{eq:2})-(\ref{eq:4}) are related to
the usual notations $f$ and $f^\dagger$ as
$f={\bar f}e^{i\phi}$ and
$f^\dagger={\bar f}^\dagger e^{-i\phi}$.

Instead of solving Eqs.(\ref{eq:2})-(\ref{eq:4}),
it is more convenient to use the following parameterization
devised by Schopohl and Maki,\cite{sch}
\begin{equation}
{\bar f} = {2 {\bar a} \over 1+{\bar a}{\bar b}} , \quad
{\bar f}^\dagger = {2 {\bar b} \over 1+{\bar a}{\bar b}} .
\label{eq:5}
\end{equation}
    From Eq.(\ref{eq:4}), $g$ is given by
\begin{equation}
g={1-{\bar a}{\bar b} \over 1+{\bar a}{\bar b} } .
\label{eq:6}
\end{equation}
    Substituting Eqs.(\ref{eq:5}) and (\ref{eq:6}) into
Eqs.(\ref{eq:2}) and (\ref{eq:3}),
we obtain the Riccati equation for $\bar a$:
\begin{equation}
\partial_\parallel {\bar a}(\omega_n,\theta,{\bf r})
-\bar\Delta(\theta,{\bf r})
+\Bigl\{2 \omega_n + i\partial_\parallel \phi
+ \bar\Delta(\theta,{\bf r}) {\bar a}(\omega_n,\theta,{\bf r})
\Bigr\}{\bar a}(\omega_n,\theta,{\bf r})=0.
\label{eq:7}
\end{equation}

\noindent
The other unknown quantity ${\bar b}$ is related to
${\bar a}$ by symmetry:
$\bar b(r_\parallel)=\bar a(-r_\parallel)$.
Far from the vortex core,
${\bar a}$ is reduced to the value of a spatially homogeneous
situation without magnetic field:
\begin{equation}
\bar a_\infty(\theta)
={\sqrt{\omega_n^2 + |\bar\Delta|^2} -\omega_n \over
\bar\Delta} \qquad (\omega_n>0).
\label{eq:8}
\end{equation}
To obtain the quasi-classical Green functions,
we integrate Eq.(\ref{eq:7}) along the trajectory
where $r_\perp$ is held constant,
using Eq.(\ref{eq:8}) as the initial value.

Second, we calculate the pair potential from
the resulting quasi-classical Green
functions by the self-consistent condition
\begin{equation}
\bar\Delta(\theta,{\bf r})=N_0
2 \pi T \sum_{\omega_n>0} \int_0^{2\pi}{d\theta \over 2\pi}
V(\theta,\theta'){\bar f}(\omega_n,\theta',{\bf r}) ,
\label{eq:9}
\end{equation}
where $N_0$ is the density of states at the Fermi surface.
The pair potential and the pairing interaction are
decomposed into $s$-, $d_{x^2-y^2}$- and $d_{xy}$-wave
components,
\begin{equation}
V({\bf \theta,\theta'})
=V_s + V_{x^2-y^2} \cos(2\theta)\cos(2\theta')
+ V_{xy} \sin(2\theta)\sin(2\theta') ,
\label{eq:10}
\end{equation}
\begin{equation}
\bar\Delta(\theta,{\bf r})=\bar\Delta_s({\bf r})
+ \bar\Delta_{x^2-y^2}({\bf r})\cos(2\theta)
+ \bar\Delta_{xy}({\bf r})\sin(2\theta)  .
\label{eq:11}
\end{equation}
    Substituting Eqs.(\ref{eq:10}) and (\ref{eq:11})
into Eq.(\ref{eq:9}), we obtain
\begin{equation}
\bar\Delta_s({\bf r})=V_s N_0
2 \pi T \sum_{\omega_n>0} \int_0^{2\pi}{d\theta \over 2\pi}
{\bar f}(\omega_n,\theta,{\bf r}) ,
\label{eq:12}
\end{equation}
\begin{equation}
\bar\Delta_{x^2-y^2}({\bf r})=V_{x^2-y^2} N_0
2 \pi T \sum_{\omega_n>0} \int_0^{2\pi}{d\theta \over 2\pi}
{\bar f}(\omega_n,\theta,{\bf r}) \cos(2\theta) ,
\label{eq:13}
\end{equation}
\begin{equation}
\bar\Delta_{xy}({\bf r})=V_{xy}N_0
2 \pi T \sum_{\omega_n>0} \int_0^{2\pi}{d\theta \over 2\pi}
{\bar f}(\omega_n,\theta,{\bf r}) \sin(2\theta) .
\label{eq:14}
\end{equation}
    As already mentioned,
at not too low temperatures, $T/T_c \ge 0.5$,
the resulting profile of $\bar\Delta_{x^2-y^2}({\bf r})$
calculated from Eq. (\ref{eq:13}) after solving the
Eilenberger equations is almost the same as
that of Eq. (\ref{eq:1}), ensuring the self-consistency
in that temperature region.

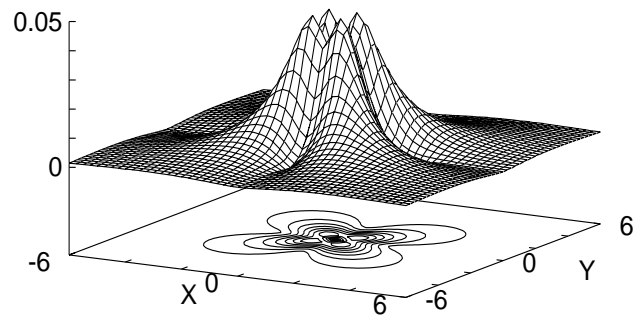
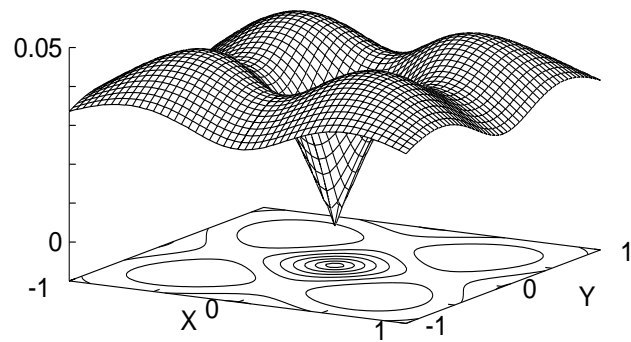
\begin{figure}
\caption{
    (a)
    Amplitude of the $s$-wave component,
$|\bar\Delta_s({\bf r})|/(V_s/V_{x^2-y^2})$,
induced around a vortex in $d_{x^2-y^2}$-wave superconductors
at $T/T_c=0.5$.
    It has the shape of a four-lobe clover.
    (b)
    The core region of (a) is focused.
    Near the center of vortex, $|\Delta_s| \propto r$.
}
\label{fig:1}
\end{figure}

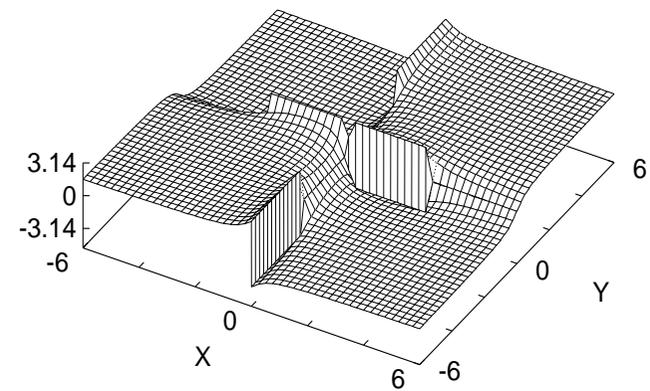
\begin{figure}
\caption{
    Phase of the induced $s$-wave component,
$\arg\bar\Delta_s({\bf r})$.
    Far from the vortex core,
the term with $e^{2i\phi}$ is dominant.
    Near the center of vortex,
the term with $e^{-2i\phi}$ is dominant.
}
\label{fig:2}
\end{figure}

    As seen from Eqs.(\ref{eq:12}) and (\ref{eq:14}),
the Green function $\bar f$ solved under
the given pair potential immediately yields the
induced $s$-wave (or $d_{xy}$-wave) component.
Figures \ref{fig:1} and \ref{fig:2} show
the $s$-wave component $\bar\Delta_s({\bf r})$
induced around a vortex in $d_{x^2-y^2}$-wave superconductors.
    As seen from Fig. \ref{fig:1}, the amplitude has
the shape of a four-lobe clover.
    Near the vortex center $|\bar\Delta_s| \propto r$ and
far from the vortex core $|\bar\Delta_s| \propto r^{-2}$.
    As shown in Fig. \ref{fig:2},
the term with $e^{-2i\phi}$ is dominant
near the vortex center and the term with $e^{2i\phi}$ is
dominant far from the vertex core.
    The induced $s$-wave component, therefore, can be written as
\begin{equation}
\Delta_s({\bf r})=\bar\Delta_s({\bf r})e^{i\phi}=
\Bigl( c_1(r)e^{-2i\phi}+c_2(r)e^{2i\phi} \Bigr) e^{i\phi},
\label{eq:15}
\end{equation}
where $c_1(r)$ and $c_2(r)$ are factors depending on $r$.
    Near the center of vortex $|c_1| > |c_2|$,
and far from the vortex core $|c_1| < |c_2|$.
    These are consistent with the results given by
Ren, Xu and Ting based on the GL theory.\cite{ren,ber}
    On lowering temperature, the amplitude
$|\bar\Delta_s({\bf r})|$ increases and the inner area
where $e^{-2i\phi}$ is dominant spreads out.

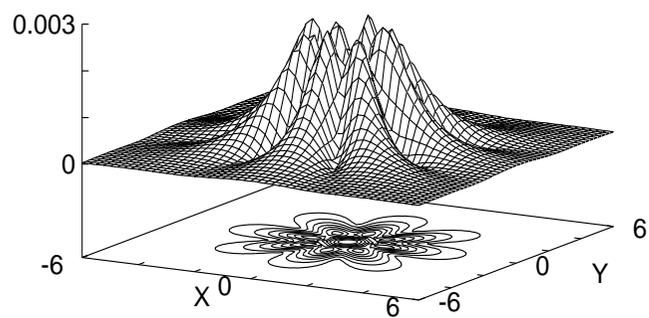
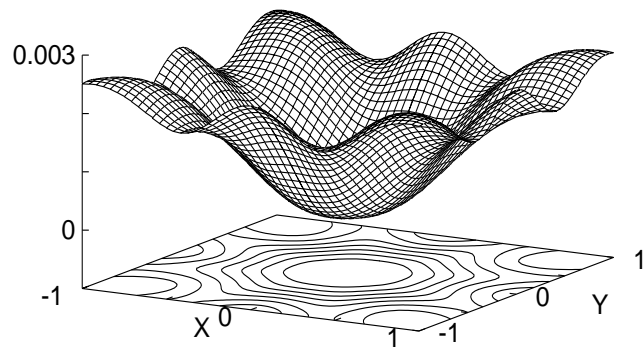
\begin{figure}
\caption{
    (a)
    Amplitude of the $d_{xy}$-wave component,
$|\bar\Delta_{xy}({\bf r})|/(V_{xy}/V_{x^2-y^2})$,
induced around a vortex in $d_{x^2-y^2}$-wave superconductors
at $T/T_c=0.5$.
    It has the shape of an octofoil.
    (b)
    The core region of (a) is focused.
    Near the center of vortex, $|\Delta_{xy}| \propto r^3$.
}
\label{fig:3}
\end{figure}

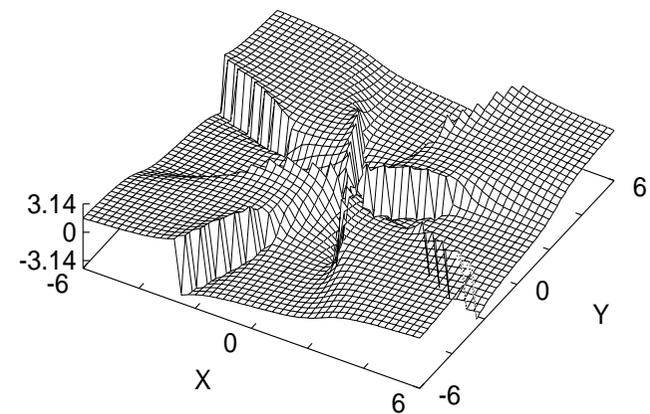
\begin{figure}
\caption{
    Phase of the induced $d_{xy}$-wave component,
$\arg\bar\Delta_{xy}({\bf r})$.
    Far from the vortex core,
the term with $e^{4i\phi}$ is dominant.
    Near the center of vortex,
the term with $e^{-4i\phi}$ is dominant.
}
\label{fig:4}
\end{figure}

    Figures \ref{fig:3} and \ref{fig:4} show
the $d_{xy}$-wave component $\bar\Delta_{xy}({\bf r})$
induced around a vortex in $d_{x^2-y^2}$-wave superconductors.
    As shown in Fig. \ref{fig:3},
the amplitude has the shape of an octofoil.
    Near the vortex center $|\bar\Delta_s| \propto r^3$ and
far from the vortex core $|\bar\Delta_s| \propto r^{-4}$.
    As seen from Fig. \ref{fig:4},
the term with $e^{-4i\phi}$ is dominant
near the vortex center and the term with $e^{4i\phi}$ is
dominant far from the vortex core.
    The induced $d_{xy}$-wave component, therefore, can be
written as
\begin{equation}
\Delta_{xy}({\bf r})=\bar\Delta_{xy}({\bf r})e^{i\phi}=
\Bigl( d_1(r)e^{-4i\phi}+d_2(r)e^{4i\phi} \Bigr) e^{i\phi},
\label{eq:16}
\end{equation}
where $d_1(r)$ and $d_2(r)$ are factors depending on $r$.
    Near the center of vortex $|d_1| > |d_2|$,
and far from the vortex core $|d_1| < |d_2|$.
    Our results of the induced $d_{xy}$-component are
also explained by the GL theory,
if the non-local correction terms are included.
    As far as the gradient terms are concerned,
the mixing of $d_{x^2-y^2}$- and $d_{xy}$-components is
absent in the usual second derivative terms
and first appears in the fourth-order derivative terms.

Let us now interpret our microscopic calculations so far
in terms of the GL framework for ease of the understanding.
The GL equations are derived from the Gor'kov equations
by the same calculation as in the case of $s$- and
$d_{x^2-y^2}$-components. \cite{ren}
    For $d_{x^2-y^2}$- and $d_{xy}$-components which are
simultaneously non-vanishing, we obtain
\widetext
\begin{eqnarray}
\alpha_{xy}\Delta_{xy}
- (\partial_x^2 +\partial_y^2)\Delta_{xy}
-\gamma \biggl\{ \biggl(
{5 \over 8}(\partial_x^2 +\partial_y^2)^2
           +\partial_x^2 \partial_y^2 \biggr)\Delta_{xy}
+ {1 \over 2}\partial_x \partial_y
            (\partial_x^2 -\partial_y^2)\Delta_{x^2-y^2} \biggr\}
\nonumber \\
+|\Delta_{xy}|^2 \Delta_{xy}
+{2 \over 3}|\Delta_{x^2-y^2}|^2 \Delta_{xy}
+{1 \over 3}\Delta_{x^2-y^2}^2 \Delta_{xy}^* =0 ,
\label{eq:17}
\end{eqnarray}
\begin{eqnarray}
-\ln\left({T_c \over T}\right) \Delta_{x^2-y^2}
- (\partial_x^2 +\partial_y^2)\Delta_{x^2-y^2}
-\gamma \biggl\{ \biggl(
{7 \over 8}(\partial_x^2 +\partial_y^2)^2
           -\partial_x^2 \partial_y^2 \biggr)\Delta_{x^2-y^2}
+{1 \over 2}\partial_x \partial_y
           (\partial_x^2 -\partial_y^2)\Delta_{xy} \biggr\}
\nonumber \\
+|\Delta_{x^2-y^2}|^2 \Delta_{x^2-y^2}
+{2 \over 3}|\Delta_{xy}|^2 \Delta_{x^2-y^2}
+{1 \over 3}\Delta_{xy}^2 \Delta_{x^2-y^2}^* =0
\label{eq:18}
\end{eqnarray}
\narrowtext\noindent
in the dimensionless form, where
$\gamma=62\zeta(5)/49\zeta(3)^2$,
$\alpha_{xy}=2\{(V_{xy}N_0)^{-1}-(V_{x^2-y^2}N_0)^{-1} \}$.
    Far from the vortex core,
we can assume $\Delta_{x^2-y^2}=\Delta_\infty e^{i\phi}$,
and the leading order terms of the GL equations (\ref{eq:17})
and (\ref{eq:18}) are written as
\begin{equation}
(\alpha_{xy}+{2 \Delta_\infty^2 \over 3})\Delta_{xy}
+ {\Delta_\infty^2 \over 3}e^{2i\phi}\Delta^*_{xy}
- {\gamma\Delta_\infty \over 2} \partial_x \partial_y
           (\partial_x^2 -\partial_y^2) e^{i\phi}=0 ,
\label{eq:19}
\end{equation}
\begin{equation}
-\ln(T_c/T)\Delta_\infty+\Delta_\infty^3=0 .
\label{eq:20}
\end{equation}
    Solving Eqs. (\ref{eq:19}) and (\ref{eq:20}), we obtain
\begin{equation}
\Delta_{xy}({\bf r})
=-{15i \over 16 r^4}\Bigl( d'_1e^{-4i\phi}
+d'_2e^{4i\phi} \Bigr) e^{i\phi},
\label{eq:21}
\end{equation}
where
$d'_1 = d' (\alpha_{xy}+3\Delta_\infty^2)$,
$d'_2 = d' (7\alpha_{xy}+5\Delta_\infty^2)$,
$d'=\{ (\alpha_{xy}+{2 \over 3}\Delta_\infty^2 )^2
-({1 \over 3}\Delta_\infty^2 )^2 \}^{-1}
\gamma \Delta_\infty $
and $\Delta_\infty=\{ \ln(T_c/T) \}^{1/2}$.
    Since $|d'_2|>|d'_1|$,
the term with $e^{4i\phi}$ is dominant.
    Near the center of vortex, the leading order of
Eqs. (\ref{eq:17}) and (\ref{eq:18}) are second
and fourth order derivative terms,
and the general solution of the GL equations
is given as follows.
    For a $d_{x^2-y^2}$-wave component,
$\Delta_{x^2-y^2}=(a_0 r + b_0 r^3 + O(r^5) )e^{i\phi}$,
where $a_0$ and $b_0$ are constants.
    For a $d_{xy}$-wave component,
assuming the form of Eq. (\ref{eq:16}) for $ \Delta_{xy} $,
we obtain $d_1=d_0 r^3 + O(r^5)$ and $d_2= O(r^5)$,
where $d_0$ is a constant.
    Therefore, the term with $e^{-4i\phi}$ is dominant
and $|\Delta_{xy}|\propto r^3$.
    These results are consistent with our
quasi-classical calculations.

    Our numerical results are quantitatively valid
for small $|V_s/V_{x^2-y^2}|$ and $|V_{xy}/V_{x^2-y^2}|$,
and for not too low temperatures.
    From Eqs. (\ref{eq:12}) and (\ref{eq:14}),
$\bar\Delta_s({\bf r})$ and $\bar\Delta_{xy}({\bf r})$
are proportional to $V_s$ and $V_{xy}$, respectively.
    If $|V_s/V_{x^2-y^2}|, |V_{xy}/V_{x^2-y^2}| \ll 1$,
$\bar\Delta_s({\bf r})$ and $\bar\Delta_{xy}({\bf r})$
are  negligibly small.
    The $s$- and $d_{xy}$-wave components, then,
do not affect $d_{x^2-y^2}$-wave superconductivity.
    In the case that $V_s$ or $V_{xy}$ is
comparable to $V_{x^2-y^2}$,
self-consistent calculations including
$s$- and $d_{xy}$-wave components are needed.
It should be noticed that at low temperatures, a
self-consistent calculation for
$\bar\Delta_{x^2-y^2}({\bf r})$ is needed since
the pair potential may deviate from Eq. (\ref{eq:1}).

    A vortex in $d_{x^2-y^2}$-wave superconductors
has a four-folded symmetric structure by the mixing of
the $s$- or $d_{xy}$-wave component.
    However, this anisotropy clearly appears only when
$V_s$ or $V_{xy}$ is comparable to $V_{x^2-y^2}$.
    It may not be probable that this condition is
satisfied in high-$T_c$ superconductors.
    Lastly, we note that
the cylindrical symmetry around a vortex line is
spontaneously broken in $d_{x^2-y^2}$-wave superconductors
even when $V_s$ and $V_{xy}$ are absent.
    The four-folded symmetric structure of $d_{x^2-y^2}$-wave
vortex becomes clear on lowering temperature,
which is confirmed by our quasi-classical calculations.\cite{ich}

    We would like to thank Professor T. Ohmi
for valuable discussions.

\begin{references}
\bibitem{self}
A full self-consistent calculation is under way, which we
believe, does not alter the essential features obtained here.
In fact, for $T/T_c \ge 0.5$ the present calculations are almost
self-consistent.
\end{references}
\end{document}